\documentclass[aps,preprint,groupedaddress]{revtex4}

\newcommand{\be}{\begin{equation}}
\newcommand{\ee}{\end{equation}}
\newcommand{\bea}{\begin{eqnarray}}
\newcommand{\eea}{\end{eqnarray}}

\begin{document}
% You should use BibTeX and revtex.bst for references
\bibliographystyle{apsrev}

% Use the \preprint command to place your local institutional report
% number on the title page in preprint mode.
% Multiple \preprint commands are allowed.
 \preprint{UAB-FT-513}

%Title of paper
\title{SUMMING THE DERIVATIVE EXPANSION OF THE EFFECTIVE ACTION}
% Optional argument for running titles on pages
%\title[]{}

% repeat the \author .. \affiliation  etc. as needed
% \email, \thanks, \homepage, \altaffiliation all apply to the current
% author. Explanatory text should go in the []'s, actual e-mail
% address or url should go in the {}'s for \email and \homepage.
% Please use the appropriate macro for the type of information

% \affiliation command applies to all authors since the last
% \affiliation command. The \affiliation command MUST follow the
% other information

\author{Eduard Mass{\'o}} 
\email[]{masso@ifae.es}
\author{Francesc Rota}
\email[]{rota@ifae.es}

%\homepage[]{Your web page}
%\thanks{}
%\altaffiliation{}
\affiliation{Grup de F{\'\i}sica Te{\`o}rica and Institut 
de F{\'\i}sica d'Altes
Energies\\Universitat Aut{\`o}noma de Barcelona\\ 
08193 Bellaterra, Barcelona, Spain}

%Collaboration name if desired (requires use of superscriptaddress
%option in \documentclass). \noaffiliation is required (may also be
%used with the \author command).
%\collaboration{}
%\noaffiliation

\date{\today}

\begin{abstract}
The derivative expansion of the effective action is a perturbative development in derivatives of the fields. The expansion breaks down when some of the derivatives are too large. We show how to sum exactly the first and second derivatives and treat perturbatively derivatives higher than second.
\end{abstract}
% insert suggested PACS numbers in braces on next line
%\pacs{}

%\maketitle must follow title, authors, abstract and PACS
\maketitle

% body of paper here - Use proper section commands
% References should be done using the \cite, \ref, and \label commands

\section{Introduction}
The effective action incorporates the effects of closed loops in a quantum field theory. This object contains all the information relevant for the low-energy effective limit. However, in general it cannot be evaluated exactly, and one has to rely on some sort of approximation.

Since the effective action is the generating functional of the one-particle-irreducible diagrams, we can develop it quite naturally in powers of the momenta flowing into the diagram vertices. This is equivalent to expand the effective Lagrangian in powers of the derivatives of the background field, an approximation that is known as the derivative expansion. 

The leading term of this derivative expansion is the effective potential, i.e., the term with no derivatives that corresponds to a constant background field case. The next orders in the expansion are obtained by developing the effective Lagrangian about this constant field case. The development is perturbative and, as such, is valid only when the field is slowly varying.

There might be instances where all the field derivatives are small except for some of them. Then, the derivative expansion breaks down and cannot be used. We have studied how to rescue such a case. To be precise, we have found a method to sum exactly the first and the second derivatives in the derivative expansion, with the higher derivatives (third, etc.) treated perturbatively. We have used a simple scalar theory to explain the details of our work, although it can be extended to more realistic theories, like QED.

Our method is developed in momentum-space. We concentrate our efforts to calculate the momentum-space Green's function $G(p)$ since, as we will see, the knowledge of $G(p)$ allows to calculate the effective Lagrangian ${\cal L}_{eff}$ quite trivially. We reobtain the derivative expansion in a very simple way. We then show how to determine an expression for $G(p)$ and ${\cal L}_{eff}$ that is exact in first and second derivatives and perturbative in higher derivatives.

\section{The effective action}
\noindent We start with the action for a scalar field
\be
S[\phi] = \int d^4x~\{\frac{1}{2}(\partial \phi)^2 - \frac{1}{2} m^2 \phi^2 - V(\phi)\}
\ee
It can be expanded around a classical background field $\phi_c$,
\be
\phi(x)=\phi_c(x)+\omega(x)
\ee
where $\omega(x)$ is a quantum field, and we obtain
\be
S[\phi]=S[\phi_c]+\Delta S[\phi_c,\omega]+\ldots
\label{Sexp}
\ee
where $\Delta S$ is defined as the piece that contains the terms bilinear in $\omega$
\be
\Delta S[\phi_c,\omega]=\int d^4x~ \frac{1}{2}\omega(x)[-\partial^2-m^2-V''(\phi_c)]\omega(x)
\ee
and the dots in (\ref{Sexp}) correspond to higher powers in $\omega$.

At the one-loop level, the effective action $S_{eff}$ is defined after integrating out the $\omega$-field
\be
e^{\frac{i}{\hbar} S_{eff}[\phi_c]}=N \int D[\omega]~e^{\frac{i}{\hbar} \Delta S[\phi_c,\omega]}
\label{Seff}
\ee
($N$ is a normalization constant).

As we said, we will apply momentum-space methods to calculate the effective action. We follow \cite{Brown} and first differentiate (\ref{Seff}) by $m^2$
\bea
\frac{i}{\hbar}\frac{\partial}{\partial m^2}S_{eff} &=& 
-\frac{i}{2\hbar}\frac{\int D[\omega]~ \omega^2 e^{\frac{i}{\hbar} \Delta S}}{\int D[\omega]~ e^{\frac{i}{\hbar} \Delta S}} \nonumber \\
&=& -\frac{1}{2}\int d^4x~ G(x,x) \nonumber \\
&=& -\frac{1}{2}\int d^4x~ \int \frac {d^4p}{(2\pi)^4}~G(p)
\label{Sdiff}
\eea
Here $G(x,x')$ is the Green's function satisfying
\be
[\partial_x^2+m^2+V''(\phi_c(x))]G(x,x')=\delta^4(x,x')
\label{G(x)}
\ee
 and $G(p)$ is its Fourier-transformed
\be
G(p)=\int d^4x~ e^{ip(x-x')}G(x,x')
\ee
The effective action and Lagrangian are formally obtained integrating the relation (\ref{Sdiff})
\bea
S_{eff}&=&\int d^4x~ {\cal L}_{eff} \nonumber \\
{\cal L}_{eff}&=& \frac{i\hbar}{2}\int dm^2 \int \frac{d^4p}{(2\pi)^4}~G(p)
\label{Leff}
\eea

In the following section we will work out the equation for $G(p)$ and its perturbative solution. From (\ref{Leff}), it is clear that the knowledge of $G(p)$ leads easily to the effective action $S_{eff}$. In general, one has constants of integration in (\ref{Leff}) which are determined by demanding $S_{eff}\rightarrow 0$ when $\phi_c \rightarrow 0$.

\section{The derivative expansion}
\noindent We will expand $V''(\phi_c)$ in (\ref{G(x)}) around a reference point $x_o$
\bea
V''(\phi_c(x))&=& V''(\phi_c(x_o))~+~(\partial_{\mu}V'')_o~(x-x_o)^{\mu}~+~\frac{1}{2}(\partial_{\mu}\partial_{\nu}V'')_o~(x-x_o)^{\mu}~(x-x_o)^{\nu} \nonumber \\
&& +~ \frac{1}{3!}(\partial_{\mu}\partial_{\nu}\partial_{\sigma}V'')_o~(x-x_o)^{\mu}~(x-x_o)^{\nu}~(x-x_o)^{\sigma}~+~...
\eea
where
\be
(\partial_{\mu}V'')_o=\left.\partial_{\mu}V''(\phi_c(x))\right|_{x=x_o}
\ee
etc.

\noindent In momentum space
\be
(x-x_o)^{\mu}\rightarrow -i\frac{\partial}{\partial p_{\mu}}
\ee
So that the equation for $G(p)$ is
\be
(-p^2+\alpha+K)~G(p)=1
\label{G(p)}
\ee
where 
\be
\alpha = m^2 + V''(\phi_c(x_o))
\label{alpha}
\ee
 and the field derivatives are in the operator $K$ :
\bea
K&=&-~i(\partial_{\mu}V'')_o~\frac{\partial}{\partial p_{\mu}} \nonumber \\
&& -~\frac{1}{2}~(\partial_{\mu}\partial_{\nu}V'')_o~\frac{\partial}{\partial p_{\mu}}\frac{\partial}{\partial p_{\nu}} \nonumber \\
&& +~i\frac{1}{3!}(\partial_{\mu}\partial_{\nu}\partial_{\sigma}V'')_o~\frac{\partial}{\partial p_{\mu}}\frac{\partial}{\partial p_{\nu}}\frac{\partial}{\partial p_{\sigma}} \nonumber \\
&& + ...
\label{defK}
\eea
When $K=0$, one can solve trivially for $G_o=G$
\be
(-p^2+\alpha)~G_o(p)=1
\ee
\be
G_o(p)=\frac{1}{-p^2+\alpha}
\label{G0}
\ee

At this point we could find the effective action in the case that the scalar field is constant, i.e., when all the derivatives are zero. Thus we would find ${\cal L}_{eff}$, and therefore the effective potential, introducing the Green's function (\ref{G0}) in (\ref{Leff}). We shall not recalculate the effective potential since it is a very well-known object; instead, we will next work out the case $K \neq 0$.

In this general case, $K\neq 0$, having a solution for $G(p)$ is equivalent to determine the inverse of the operator acting on $G(p)$ in (\ref{G(p)}), namely,
\be
\frac{1}{-p^2+\alpha+K}
\ee
We now use the following expansion
\be
\frac{1}{-p^2+\alpha+K} = \frac{1}{-p^2+\alpha}-\frac{1}{-p^2+\alpha}K\frac{1}{-p^2+\alpha}+\frac{1}{-p^2+\alpha}K\frac{1}{-p^2+\alpha}K\frac{1}{-p^2+\alpha}-\ldots 
\ee
so that
\bea
G(p)&=& G_o(p)~-~G_o(p)KG_o(p)~+~G_o(p)KG_o(p)KG_o(p)~+~\ldots \nonumber \\
&=& G_o(p)\sum_{m=0}^{\infty}(-1)^m(K G_o(p))^m
\label{DerExp}
\eea

This is an infinite expansion of $G(p)$ in powers of derivatives. With (\ref{DerExp}) we can calculate $G(p)$ up to any desired order and then integrate it as we indicate in (\ref{Leff}) and obtain the effective Lagrangian ${\cal L}_{eff}$.

This development of the effective Lagrangian is called the derivative expansion and was first obtained by \cite{Chan} and by \cite{Fraser} (see also \cite{Cheyette}, \cite{Fraser2}). We think our method to derive this expansion is quite simple.

\section{Summing the derivative expansion}
The derivative expansion is a perturbative expansion in the field derivatives. As such, it may be useful when all the derivatives are small, i.e.,
\be
\frac{1}{\alpha^{1+n/2}}\,\frac{\partial^n V''_o}{\partial x^{\mu_1}\,\partial x^{\mu_2}\cdots\partial x^{\mu_n}} \ll 1 
\ee
$n=1,2,\ldots$ 

Actually, there has been much discussion in the literature about the convergence of the derivative expansions \cite{Perry,Li,Chan2,Dunne,Dunne2}. Of course it would be interesting to sum exactly some of the derivative terms. We know that one cannot sum all the derivative expansion for an arbitrary theory, so we have studied which part of the expansion we can sum. The conclusion of our study is that we are able to sum exactly the terms in $\alpha$, eq.(\ref{alpha}), and in the first and second derivatives
\bea
\beta_{\mu} &=& (\partial_{\mu}V'')_o \nonumber \\
\gamma^2_{\mu\nu} &=& 2(\partial_{\mu}\partial_{\nu}V'')_o
\label{gamma}
\eea
and leave the other terms, i.e., the higher-than-second derivatives, as a perturbation expansion. In the rest of the paper we will show how to do it.

Let us start discussing first the case that $K$ in (\ref{defK}) has only first and second derivatives,
\be
K=-i\beta_{\mu} \frac{\partial}{\partial p_{\mu}}-\frac{1}{4}\gamma^2_{\mu\nu}\frac{\partial^2}{\partial p_{\mu}\partial p_{\nu}} \equiv K_2
\ee
so that our eq.(\ref{G(p)}) reads
\be
(-p^2+\alpha+K_2)~G_2(p)=1
\label{G2}
\ee

The finding that (\ref{G2}) has an exact solution goes back to the work of Schwinger \cite{Schwinger}, but we will follow the momentum-space methods of \cite{Brown}. We will seek solutions of the type
\be
G(p)=\int_0^\infty ds~e^{F(s)}e^{-\alpha s}
\label{Gtp}
\ee
($s$ is called proper time). When this ansatz is introduced in eq.(\ref{G(p)}), we end up with
\be
\int_0^\infty ds~H(s) e^{F(s)}e^{-\alpha s}=1
\label{eqH}
\ee
where
\be
H(s) e^{F(s)}=(-p^2+\alpha+K)e^{F(s)}
\label{eqH2}
\ee
If a solution to (\ref{eqH}) exists, it fulfills
\be
H(s)=-\frac{\partial}{\partial s}(F-\alpha s)
\label{solH}
\ee
with boundary conditions
\bea
F-s\alpha \rightarrow 0 ~~~&\mathrm{when}&~~~ s \rightarrow 0 \nonumber \\
F-s\alpha \rightarrow -\infty ~~~ &\mathrm{when}& ~~~ s \rightarrow \infty
\label{bc}
\eea
When $K=0$, the ansatz (\ref{Gtp}) for $G_o$ corresponds to
\be
F=s~p^2
\label{FreeF}
\ee
clearly, then we have
\be
G_o=\int_0^\infty ds~ e^{(p^2-\alpha)s} = \frac{1}{-p^2+\alpha}
\ee
in agreement with (\ref{G0}).

\noindent When $K=K_2$, we follow reference \cite{Brown} and generalize (\ref{FreeF})
\be
F=A_{2\mu\nu}~p^{\mu}p^{\nu} + B_{2\mu}~p^{\mu} + C_2 \equiv F_2
\label{F2}
\ee
where $A_2$, $B_2$, $C_2$ are functions of $s$ to be determined. The function $H(s)$ in (\ref{eqH}) is easily calculated 
\bea
H&=&-p^2+\alpha-i\beta_{\mu}B_2^{\mu}-\frac{1}{2}\gamma^2_{\mu\nu}A_2^{\mu\nu}-\frac{1}{4}\gamma^2_{\mu\nu}B_2^{\mu}B_2^{\nu}-\left[2i\beta_{\nu}A_2^{\nu\mu}+\gamma^2_{\nu\rho}B_2^{\nu}A_2^{\rho\mu}\right]p_{\mu} -\gamma^2_{\rho\sigma}A_2^{\rho\mu}A_2^{\sigma\nu}p_{\mu}p_{\nu} \nonumber \\ &\equiv& H_2
\eea
Both $F_2$ and $H_2$ are polynomials of second order in $p$ and thus we can use (\ref{solH}) and equate equal powers of $p$ on both sides. We obtain the following differential equations
\bea
&&\frac{\partial}{\partial s}A_{2\mu\nu} = g_{\mu\nu}+\gamma^2_{\rho\sigma}A_{2\mu}^{\rho}A_{2\nu}^{\sigma} \nonumber \\
&&\frac{\partial}{\partial s}B_{2\mu} = 2i\beta_{\nu}A_{2\mu}^{\nu}+\gamma^2_{\rho\nu}B_2^{\rho}A_{2\mu}^{\nu} \nonumber \\
&&\frac{\partial}{\partial s}C_2 = i\beta_{\mu}B_2^{\mu}+\frac{1}{2}\gamma^2_{\mu\nu}A_2^{\mu\nu}+\frac{1}{4}\gamma^2_{\mu\nu}B_2^{\mu}B_2^{\nu} 
\label{SistDif0}
\eea
Taking into account the boundary conditions (\ref{bc}), the solutions are
\bea
A_{2\mu\nu} &=& \gamma^{-1}_{\mu\rho} (\tan \gamma s)^{\rho}_{\nu} \nonumber \\
B_{2\mu} &=& -2i \gamma^{-2}_{\mu\rho} [g^{\rho}_{\sigma}-(\sec \gamma s)^{\rho}_{\sigma}] \beta^{\sigma} \nonumber \\
C_2 &=& -\frac{1}{2}~ tr \ln (\cos \gamma s)-\beta^{\mu}\gamma^{-3}_{\mu\rho}(\tan \gamma s - \gamma s)^{\rho}_{\nu}\beta^{\nu}
\label{SolSistDif0}
\eea
Putting these solutions in (\ref{F2}), this determines
\be
G_2(p) = \int^{\infty}_0 ds~e^{F_2}e^{-\alpha s} 
\ee

This expression for $G_2$ is valid for arbitrary values of the first and second derivatives $\beta$ and $\gamma$. From $G_2$ one can get, using (\ref{Leff}), the effective Lagrangian that we call ${\cal L}_2$ (see below eq. (\ref{L2})).

When higher derivatives are considered there is no longer an exact solution for $G(p)$ and ${\cal L}_{eff}$. However, we will show that we can get a solution valid for arbitrary $\alpha$, $\beta$, and $\gamma$ and perturbative in higher derivatives. For the sake of clarity, let us explicitly demonstrate this in the case of having a third derivative :
\be
K = -i\beta_{\mu} \frac{\partial}{\partial p_{\mu}}-\frac{1}{4}\gamma^2_{\mu\nu}\frac{\partial^2}{\partial p_{\mu}\partial p_{\nu}} +i~g~\delta_{\mu\nu\rho}\frac{\partial^3}{\partial p_{\mu}\partial p_{\nu}\partial p_{\rho}}
\ee
with $g\delta_{\mu\nu\rho} = \left.\frac{1}{3!}\partial_{\mu}\partial_{\nu}\partial_{\rho} V'' \right|_o$.

We adopt the ansatz (\ref{Gtp}) with
\be
F = A_{\mu\nu}~p^{\mu}p^{\nu} + B_{\mu}~p^{\mu} + C + g~D_{\mu\nu\rho}~p^{\mu}p^{\nu}p^{\rho} 
\ee
where $A$, $B$, $C$ and $D$ are $s$-functions to be determined. When the ansatz is introduced in (\ref{G(p)}) we get equation (\ref{eqH}) with
\bea
H &=& -p^2+\alpha-i\beta_{\alpha}B^{\alpha}-\frac{1}{4}\gamma^2_{\alpha\beta}(B^{\alpha}B^{\beta}+2A^{\alpha\beta})-(\gamma^2_{\alpha\beta}B^{\beta}+2i\beta_{\alpha})A^{\alpha\mu}p_{\mu} - \gamma^2_{\alpha\beta}A^{\alpha\mu}A^{\beta\nu}p_{\mu}p_{\nu} \nonumber \\
& +g & \left\{ i~\delta_{\alpha\beta\gamma}B^{\alpha}(B^{\beta}B^{\gamma}+6A^{\beta\gamma})- \left[ \frac{3}{2} \gamma^2_{\alpha\beta}D^{\alpha\beta\mu}-6i~\delta_{\alpha\beta\gamma}A^{\gamma\mu}(B^{\alpha}B^{\beta}+2A^{\alpha\beta}) \right]p_{\mu} - \right. \nonumber \\
&& \, \left[\frac{3}{2}(\gamma^2_{\alpha\beta}B^{\beta}+2i\beta_{\alpha})D^{\alpha\mu\nu} - 12i~\delta_{\alpha\beta\gamma}B^{\alpha}A^{\beta\mu}A^{\gamma\nu} \right]p_{\mu}p_{\nu} - \nonumber \\ 
&& \left. \left[3\gamma^2_{\alpha\beta}A^{\alpha\mu}D^{\beta\nu\rho} - 8i~\delta_{\alpha\beta\gamma}A^{\alpha\mu}A^{\beta\nu}A^{\gamma\rho} \right]p_{\mu}p_{\nu}p_{\rho}~ \right\} + O(g^2)
\label{H3}
\eea
Here we have used the fact that $A_{\mu\nu}$ and $D_{\mu\nu\rho}$ are symmetric tensors.

In order that (\ref{solH}) has a solution, it is a necessary condition that the two polynomials in $p$, $F$ and $H$, have the same degree. $F$ is of order $p^3$. However, the $O(g^2)$ term in $H$ in (\ref{H3}) is of order $p^6$, and this is why we are not able to obtain an exact solution for all $g$. At first order in $g$, $H$ is of order $p^3$ and we can equate equal powers of $p$ in eq.(\ref{solH}). In this way, we are able to find equations for $A$, $B$, $C$, and $D$. It is convenient to write
\bea
A &=& A_2 + g~\Delta A + O(g^2)\nonumber \\ 
B &=& B_2 + g~\Delta B + O(g^2)\nonumber \\ 
C &=& C_2 + g~\Delta C + O(g^2)
\eea
Then, using (\ref{solH}) and (\ref{SistDif0}), we find that the unknown $s$-functions $\Delta A$, $\Delta B$, $\Delta C$, and $D$ satisfy the differential equations 
\bea
 \frac{\partial}{\partial s}\Delta A_{\mu\nu} &=& 2 \gamma^2_{\alpha\beta}(A_{2\mu}^{\alpha} \Delta A^{\beta}_{\nu}+A_{2\nu}^{\alpha} \Delta A^{\beta}_{\mu}) + \frac{3}{2}(\gamma^2_{\alpha\beta}B_2^{\beta}+2i\beta_{\alpha})D^{\alpha}_{\mu\nu} \nonumber \\
&& - 12i~\delta_{\alpha\beta\gamma}B_2^{\alpha}A_{2\mu}^{\beta}A_{2\nu}^{\gamma} \label{eqA} \\
 \frac{\partial}{\partial s}\Delta B_{\mu} &=& \gamma^2_{\alpha\beta}A^{\beta}_{2\mu}\Delta B^{\alpha} +(\gamma^2_{\alpha\beta}B_2^{\beta}+2i\beta_{\alpha})\Delta A^{\alpha}_{\mu} + \frac{3}{2} \gamma^2_{\alpha\beta}D^{\alpha\beta}_{\mu} \nonumber \\
&&-6i~\delta_{\alpha\beta\gamma}A_{2\mu}^{\gamma}(B_2^{\alpha}B_2^{\beta}+2A_2^{\alpha\beta}) \label{eqB} \\
 \frac{\partial}{\partial s}\Delta C &=& \frac{1}{2}\gamma^2_{\alpha\beta}\Delta A^{\alpha\beta} + \frac{1}{2}(\gamma^2_{\alpha\beta}B_2^{\beta}+2i\beta_{\alpha})\Delta B^{\alpha} - i~\delta_{\alpha\beta\gamma}B_2^{\alpha}(B_2^{\beta}B_2^{\gamma}+6A_2^{\beta\gamma}) \label{eqC} \\
 \frac{\partial}{\partial s}D_{\mu\nu\rho} &=& 3\gamma^2_{\alpha\beta}(A^{\alpha}_{2\mu}D^{\beta}_{\nu\rho}+A^{\alpha}_{2\nu}D^{\beta}_{\rho\mu}+A^{\alpha}_{2\rho}D^{\beta}_{\mu\nu}) - 8i~\delta_{\alpha\beta\gamma}A_2^{\alpha\mu}A_2^{\beta\nu}A_2^{\gamma\rho} \label{eqD}
\eea

These are linear differential equations and can be solved, although the general solution is extremely complicated. We will discuss this issue in the next section. Here we will assume we got the solution of (\ref{eqA}-\ref{eqD}) and we will work out the form of the effective Lagrangian.

The effective Lagrangian is obtained from $G(p)$ by means of eq.(\ref{Leff}). We see we need to evaluate
\be
{\cal L}_{eff}= \frac{i\hbar}{2}\int^\infty_0 ds \int dm^2~ e^{-\alpha s} \int \frac{d^4p}{(2\pi)^4}~ e^{F(s)}
\label{Leff2}
\ee

Since we work in perturbation theory at first order in the coupling $g$, the expression of ${\cal L}_{eff}$ has to be of the form
\be
{\cal L}_{eff} = {\cal L}_2 + g~\Delta{\cal L}_3
\label{Lpert}
\ee
That this is indeed the case can be seen by expanding the integrand
\be
e^F = e^{F_2} \left[ 1 + g \left( \Delta A^{\mu\nu}p_\mu p_\nu + \Delta B^\mu p_\mu + \Delta C +  D^{\mu\nu\rho}p_\mu p_\nu p_\rho \right) + O(g^2) \right]
\ee
and inserting the expansion in (\ref{Leff2}). To integrate in momentum, we shift $p$ in such a way that $F_2$, eq.(\ref{F2}), is quadratic in the shifted momentum and the integral becomes a simple Gaussian. We also integrate in $m^2$. In this integration we obtain an integration constant that will be fixed by demanding ${\cal L}_{eff} \rightarrow 0$ when $\phi_c \rightarrow 0$. After the integration we obtain
\be
{\cal L}_2 =\frac{\hbar}{2(4\pi)^2} \int^\infty_0 ds~ \frac{1}{s^3} \left[\frac{e^{-\alpha s}\,s^2}{\sqrt{det A_2}}~ e^{-\frac{1}{4}B_2 A_2^{-1} B_2 + C_2} - e^{-m^2 s} \right]
\label{L2}
\ee  
and
\bea
\Delta{\cal L}_3 &=& \frac{\hbar}{2(4\pi)^2} \int^\infty_0 ds~ \frac{e^{-\alpha s}}{s} \frac{1}{\sqrt{det A_2}}~ e^{-\frac{1}{4}B_2 A_2^{-1} B_2 + C_2}~\Delta I \nonumber \\
\Delta I &=& \frac{1}{2} \Delta A^{\mu\nu} \left[\frac{1}{2}\left(A_2^{-1}B_2\right)_\mu \left(A_2^{-1}B_2\right)_\nu - A_{2\mu\nu}^{-1} \right] -\frac{1}{2} \Delta B^\mu \left(A_2^{-1}B_2\right)_\mu + \Delta C \nonumber \\
&& + \frac{1}{4} D^{\mu\nu\rho} \left[3 A_{2\mu\nu}^{-1}\left(A_2^{-1}B_2\right)_\rho - \frac{1}{2} \left(A_2^{-1}B_2\right)_\mu \left(A_2^{-1}B_2\right)_\nu \left(A_2^{-1}B_2\right)_\rho
 \right]
\label{Leff3}
\eea

We have determined the expression for the effective Lagrangian (\ref{Lpert}), that is valid for arbitrary $\alpha$, $\beta$, $\gamma$ and first-order in the third derivative $g\,\delta$. To ${\cal L}_2$ one should add counter terms coming from the renormalization of the mass and coupling constants in $V(\phi_c)$. These make ${\cal L}_2$ finite. 

We could obtain similar expressions when higher derivatives are considered. To show this, first consider the case that we have a $n$-th order derivatives instead of third order
\be
K = i \beta \frac{\partial}{\partial p} - \frac{1}{4} \gamma^2 \frac{\partial^2}{\partial p^2} + g~\rho \frac{\partial^n}{\partial p^n}
\label{Kn}
\ee
(in order to simplify notation we write without indices; $\rho$ would be a tensor with $n$ indices, etc.).

The ansatz that we have to consider still is eq.(\ref{Gtp}) with
\be
F = A\,p^2 + B\,p + C + g\,D_3\,p^3 + \ldots + g\,D_n\,p^n
\label{Fn}
\ee
(again no indices appear; $D_i$ is a tensor with $i$-indices, etc).

Let us discuss the order in $p$ of $H$ in (\ref{eqH}). With the action of the first and second derivatives in (\ref{eqH2}) we obtain contributions to $H$ that have maximum degree
\be
g\,p^n
\label{gpn}
\ee
at first order in $g$. There are terms of higher degree in $p$, but are $O(g^2)$ and we neglect them. When the $n$-th derivative in $K$ acts on $G$ we obtain again (\ref{gpn}) as the maximum degree. There are also higher degrees in $p$, but $O(g^2)$. Since $F$ in (\ref{Fn}) is order $p^n$, eq.(\ref{solH}) is consistent. Equating equal powers of $p$ we obtain $n+1$ linear differential equations for the functions $A$, $B$, $C$,$D_3$,$\ldots$, $D_n$. For a general tensor $\rho$ in (\ref{Kn}), to find a solution for these functions can be a painful task. Our point is that the solutions for $A$, $B$, $C$,$D_3$,$\ldots$, $D_n$ could be found and they determine $G(p)$ that leads to 
\be
{\cal L}_{eff} = {\cal L}_2 + g\,\Delta{\cal L}_n
\label{Leffn}
\ee

Now that we have discussed what happens when we have the $n$-th derivative, eq.(\ref{Kn}), it is easy to see what happens in the general case of a finite number of derivatives : first, second and then from third derivative  until $n$-th derivative. Let each of these higher than second derivatives be proportional to a coupling constant: from $g_3$ until $g_n$. Since we work at first-order in the coupling constants $g_i$, the overall modification to ${\cal L}_2$ is simply the sum of the individual modifications coming from each $g_i$. Finally we would obtain
\be
{\cal L}_{eff} = {\cal L}_2 + \sum^n_{i=3}g_i\,\Delta{\cal L}_i
\label{Leffni}
\ee
where each $\Delta{\cal L}_i$ is obtained like in (\ref{Leffn}).
The effective Lagrangian (\ref{Leffni}) is exact in $\alpha$, $\beta$, $\gamma$ and first-order in $g_3$, $\ldots$, $g_n$.

Until now we have limited  ourselves to first-order perturbation theory. Our method is really perturbative and thus we can go to higher orders in the coupling constants. Let us show this by going back to $K$ with a $n$-th derivative, eq.(\ref{Kn}), and work at second order in $g$. 

We use the ansatz (\ref{Gtp}) with 
\bea
F &=& A\,p^2 + B\,p + C +(g\,D_3 + g^2\,E_3)\,p^3 + \dots + (g\,D_n + g^2\,E_n)\,p^n \nonumber \\
&& + g^2\,D_{n+1}\,p^{n+1} + \ldots + g^2\,D_{2(n-1)}\,p^{2(n-1)}
\eea 
and
\bea
A &=& A_2 + g\,\Delta A + g^2\,\Delta\widetilde{A} + O(g^3) \nonumber \\
B &=& B_2 + g\,\Delta B + g^2\,\Delta\widetilde{B} + O(g^3)\nonumber \\
C &=& C_2 + g\,\Delta C + g^2\,\Delta\widetilde{C} + O(g^3) 
\eea

The functions $D_3$, $\ldots$, $D_n$, $\Delta A$, $\Delta B$, $\Delta C$ contain the modifications arising at first-order in $g$. Let us concentrate in the second-order corrections. It is not difficult to see that, at order $g^2$, the action of $K$ leads to terms in $H$ that have a maximum degree of
\be
g^2\,p^{2(n-1)}
\ee
and thus we have from eq.(\ref{solH}) consistent equations  for $D_{n+1}$, $\ldots$, $D_{2(n-1)}$, $E_3$,$\ldots$, $E_n$, $\Delta\widetilde{A}$, $\Delta\widetilde{B}$, $\Delta\widetilde{C}$. The final effective Lagrangian will have the form
\be
{\cal L}_{eff} = {\cal L}_2 + g\,\Delta{\cal L}_n + g^2\,\Delta\widetilde{{\cal L}}_n
\ee

The perturbative expansion could be calculated up to any order in the coupling constants corresponding to a finite number of derivatives.

\section{Explicit solutions}

The linear differential equations (\ref{eqA}-\ref{eqD}) can be solved in the standard way. It is easy to see that we have to solve them in the following order. First, eq.(\ref{eqD}) has the solutions
\be
D_{\mu\nu\rho} = -2i~\left(\frac{1}{\gamma \cos \gamma s}\right)^\lambda_\mu \left(\frac{1}{\gamma \cos \gamma s}\right)^\theta_\nu \left(\frac{1}{\gamma \cos \gamma s}\right)^\omega_\rho \, \delta_{\sigma\pi\tau} \sum_{i=0}^3 \theta_i \left[ \frac{\cos(T^is)-1}{T^i} \right]^{\sigma\pi\tau}_{\lambda\theta\omega}
\ee
where
\be
T^i_{\lambda\sigma\theta\pi\omega\tau} = a^i \, \gamma_{\lambda\sigma}\otimes g_{\theta\pi}\otimes g_{\omega\tau} +
b^i \, g_{\lambda\sigma}\otimes \gamma_{\theta\pi}\otimes g_{\omega\tau} +
c^i \, g_{\lambda\sigma}\otimes g_{\theta\pi}\otimes \gamma_{\omega\tau}
\ee
with
\bea
a^i &=& 1 \hspace{1cm} i\neq 1 , \hspace{1.5cm} a^1 = -1 \nonumber \\
b^i &=& 1 \hspace{1cm} i\neq 2 , \hspace{1.5cm} b^2 = -1 \nonumber \\
c^i &=& 1 \hspace{1cm} i\neq 3 , \hspace{1.5cm} c^3 = -1 \nonumber \\
\theta_i &=& -1 \hspace{.66cm} i\neq 0 , \hspace{1.5cm} \theta_0 = 1 
\eea

Having $D$ one can solve (\ref{eqA})  
\bea
\Delta A_{\mu\nu} &=& \left(\frac{1}{\gamma \cos \gamma s}\right)^\alpha_\mu \left(\frac{1}{\gamma \cos \gamma s}\right)^\beta_\nu \left\{ \int ds~ (\cos\gamma s)^\sigma_\alpha(\cos\gamma s)^\gamma_\beta \, M_{\sigma\gamma} + K_{\alpha\beta} \right\} \label{solA}\\
M_{\mu\nu} &=& 3i \,\left[ \beta^\alpha (\sec\gamma s)^\beta_\alpha D_{\beta\mu\nu}-4~\delta_{\alpha\beta\gamma} B_2^\alpha A^\beta_{2\mu}A^\gamma_{2\nu} \right] \nonumber
\eea

The constant of integration $K_{\alpha\beta}$ has to be chosen in such a way that the boundary conditions (\ref{bc}) are satisfied. $A_2$ and $B_2$ are given by (\ref{SolSistDif0}).

Afterwards, we can solve (\ref{eqB})
\bea
\Delta B_\mu &=& \left(\frac{1}{\gamma \cos \gamma s}\right)^\alpha_\mu \left\{ \int ds~ (\cos\gamma s)^\beta_\alpha \, V_\beta\ + K_{\alpha} \right\}  \label{solB}\\
V_\mu &=& 2i\, \beta_\alpha (\sec\gamma s)^\beta_\alpha \Delta A_{\beta\mu} + \frac{3}{2}\gamma^2_{\alpha\beta}D^{\alpha\beta}_\mu - 6i\, \delta_{\alpha\beta\gamma}A^\gamma_{2\mu}(2A_2^{\alpha\beta}+B_2^\alpha B_2^\beta) \nonumber
\eea
and finally (\ref{eqC})
\be
\Delta C = \int ds\left\{ \frac{1}{2}\gamma^2_{\alpha\beta}\Delta A^{\alpha\beta} + i~\beta^\alpha (\sec\gamma s)^\beta_\alpha \Delta B_\beta - i~\delta_{\alpha\beta\gamma}B_2^{\alpha}(B_2^{\beta}B_2^{\gamma}+6A_2^{\beta\gamma})    \right\} + K
\label{solC}
\ee
As before, $K_\alpha$ in (\ref{solB}) and $K$ in (\ref{solC}) are fixed by demanding (\ref{bc}).

We have evaluated the expressions for $D$, $\Delta A$, $\Delta B$ and $\Delta C$ for general $\delta_{\alpha\beta\gamma}$ but they are rather long and not particularly illuminating. We do not display them here. In practice, we may be interested in cases where some of the $\delta_{\alpha\beta\gamma}$ elements are zero and then the expressions for $D,\,\Delta A,\,\Delta B$ and $\Delta C$ are much easier to handle. For example, let us work out the case that $\delta_{\alpha\beta\gamma}$ is diagonal. First, we notice that we can always work in a basis where $\gamma^2_{\mu\nu}$ in (\ref{gamma}) is diagonal,
\be
\gamma^\mu_\nu = \left( 
\begin{array}{clcr}\,\gamma_o\,&&&\\&\,\gamma_1\,&&\\&&\,\gamma_2\,&\\&&&\,\gamma_3\, \end{array} \right)
\ee
Then, since $\delta_{\alpha\beta\gamma}$ is diagonal, it follows that $D_{\mu\nu\rho},\,\Delta A_{\mu\nu}$ are diagonal and given by
\bea
D_{ooo} &=& \frac{8i\,\delta_{ooo}}{3\,\gamma_o^4} \left(\frac{3}{\cos^2(s\,\gamma_o)} - \frac{2}{\cos^3(s\,\gamma_o)} - 1 \right)\\
\Delta A_{oo} &=& \frac{4\,\beta_o\,\delta_{ooo}}{\gamma_o^4}\,\left( \frac{3\,\tan(s\,\gamma_o)}{\gamma_o} - \frac{3\,s}{\cos^2 (s\,\gamma_o)} + \frac{4\,\tan(s\,\gamma_o)}{\gamma_o\cos^2(s\,\gamma_o)\,} -  \frac{4\,\tan(s\,\gamma_o)}{\gamma_o\cos(s\,\gamma_o)\,} \right)  
\eea
We have only displayed the temporal diagonal terms. To get the other elements, we have to change all the $_o$ subindices by the desired one. Although we are using the Minkowskian metric $(+,-,-,-)$ there is no change of sign.

We also have
\bea
\Delta B_o &=& \frac{8i \,\beta_o^2\delta_{ooo}}{\gamma_o^6} \left( \frac{2}{\cos^3(s\,\gamma_o)} -3 - \frac{1}{\cos^2(s\,\gamma_o)} + 
  \frac{2}{\cos(s\,\gamma_o)} - \frac{3\,s\,\gamma_o\,\tan(s\,\gamma_o)}
   {\cos(s\,\gamma_o)} \right) \nonumber \\
&+& \frac{8i \,\beta_o^2\delta_{ooo}}{\gamma_o^6} \left( \frac{\gamma_o^3\,\tan(s\,\gamma_o)} {\beta_o^2} - \frac{\gamma _o^3\,\tan(s\,\gamma_o)}
   {\beta_o^2\,\cos(s\,\gamma_o)}  \right)  \label{DeltaB} \\
\Delta C &=& \frac{4\,\beta_o\,\delta_{ooo}}{\gamma_o^4} \left( \frac{2}{\cos^2(s\,\gamma_o)} - \frac{1}{\cos(s\,\gamma_o)} + 
  \frac{2\,s\,\beta_o^2}{\gamma_o^2} + 
  \frac{3\,s\,\beta_o^2}{\gamma_o^2\,\cos^2(s\,\gamma_o)} - 
  \frac{11\,\beta_o^2\,\tan(s\,\gamma_o)}{3\,\gamma_o^3} \right. \nonumber \\
&& \left. - \frac{4\,\beta_o^2\,\tan(s\,\gamma_o)}
   {3\,\gamma_o^3\,\cos^2(s\,\gamma_o)} - 
  \frac{3\,s\,\gamma_o\,\tan(s\,\gamma_o)}{2} -1 \right) + \nonumber \\
&& \sum_{i=1}^3  \frac{4\,\beta_i\,\delta_{iii}}{\gamma_i^4} \left( \frac{1}{\cos(s\,\gamma_i)} -\frac{2}{\cos^2(s\,\gamma_i)} + 
  \frac{2\,s\,\beta_i^2}{\gamma_i^2} + 
  \frac{3\,s\,\beta_i^2}{\gamma_i^2\,\cos^2(s\,\gamma_i)} - 
  \frac{11\,\beta_i^2\,\tan(s\,\gamma_i)}{3\,\gamma_i^3} \right. \nonumber \\
&& \left. - \frac{4\,\beta_i^2\,\tan(s\,\gamma_i)}{3\,\gamma_i^3\,\cos^2(s\,\gamma_i)} + 
  \frac{3\,s\,\gamma_i\,\tan(s\,\gamma_i)}{2} + 1 \right)
\eea

As before, if we want the other elements of $\Delta B$ we have to change all the $_o$ subindices, but in this case we also have to change the sign of the second parenthesis in (\ref{DeltaB}).

With these solutions we easily get the corresponding effective Lagrangian, using eqs.(\ref{Lpert}), (\ref{L2}), and (\ref{Leff3}). We find the following result
\bea
\Delta I &=& \frac{2\,\beta_o\,\delta_{ooo}}{\gamma_o^6} \left( 4\,s\,\beta_o^2 + 6\,s\,\beta_o^2\,\sec^2 (\frac{s\,\gamma_o}{2}) - 7\,\gamma_o^2 + 
  4\,\gamma_o^2\,\sec^2 (\frac{s\,\gamma_o}{2}) + 
  \frac{3}{2}\,s\,\gamma_o^3\,\cot (\frac{s\,\gamma_o}{2}) \right. \nonumber \\
&& \left.  -  \frac{44\,\beta_o^2\,\tan (\frac{s\,\gamma_o}{2})}{3\,\gamma_o} - \frac{16\,\beta_o^2\,\sec^2 (\frac{s\,\gamma_o}{2})\,\tan (\frac{s\,\gamma_o}{2})}{3\,\gamma_o} - \frac{3}{2}\,s\,\gamma_o^3\,\tan (\frac{s\,\gamma_o}{2})  \right) + \nonumber \\
&& \sum^3_{i=1} \frac{2\,\beta_i\,\delta_{iii}}{\gamma_i^6} \left( 4\,s\,\beta_i^2 + 6\,s\,\beta_i^2\,\sec^2 (\frac{s\,\gamma_i}{2}) + 7\,\gamma_i^2 - 4\,\gamma_i^2\,\sec^2 (\frac{s\,\gamma_i}{2}) - \frac{3}{2}\,s\,\gamma_i^3\,\cot (\frac{s\,\gamma_i}{2}) \right. \nonumber \\
&& \left.  -  \frac{44\,\beta_i^2\,\tan (\frac{s\,\gamma_i}{2})}{3\,\gamma_i} - \frac{16\,\beta_i^2\,\sec^2 (\frac{s\,\gamma_i}{2})\,\tan (\frac{s\,\gamma_i}{2})}{3\,\gamma_i} + \frac{3}{2}\,s\,\gamma_i^3\,\tan (\frac{s\,\gamma_i}{2})  \right)
\eea

Another particular case that leads to a simple solution is when $\gamma^2=0$, i.e., the second derivative vanishes. In that case, from (\ref{SolSistDif0}) we have
\bea
A_{2\mu\nu} &=& s\,g_{\mu\nu} \nonumber \\
B_{2\mu} &=& i\,s\,\beta_\mu \nonumber \\
C_2 &=& -\frac{s^3}{3}\beta_\mu\beta^\mu
\eea
and then we get
\bea
D_{\mu\nu\rho} &=& -2i\,s^4 \delta_{\mu\nu\rho} \nonumber \\
\Delta A_{\mu\nu} &=& \frac{18}{5} s^5 \beta^\alpha \delta_{\alpha\mu\nu} \nonumber \\
\Delta B_\mu &=& i\,s^3 \delta_{\mu\nu\rho}\left( \frac{11}{5}s^3\beta^\nu \beta^\rho -4g^{\nu\rho}\right) \nonumber \\
\Delta C &=& \frac{5}{2}s^4 \delta^\alpha_{\alpha\mu}\beta^\mu - \frac{16}{35}s^7 \delta_{\mu\nu\rho}\beta^\mu\beta^\nu\beta^\rho
\eea

When we put these expressions in (\ref{Leff3}) and then in (\ref{Lpert}), we get the effective Lagrangian for this particular case
\be
{\cal L}_{eff} = \frac{\hbar}{32\pi^2} \int_0^\infty \frac{ds}{s^3}\left\{ e^{-\alpha s-\frac{1}{12}\beta^2 s^3} \left[1 + g \left(g^{\mu\nu}-\frac{s^3}{28}\beta^\mu\beta^\nu \right)\frac{s^4}{5}\beta^\rho\delta_{\mu\nu\rho} \right] -e^{-m^2s} \right\}
\ee

% If in twocolumn mode, this environment will move to single column
% format so that long equations can be displayed. Use
% sparingly.
% Note: this may cause bad behavior if this occurs near a page
% break - this can be worked around by adding an explicit \pagebreak.
%\begin{widetext}
% put long equation here
%\end{widetext}

\section*{Note added}
We would like to comment that it is possible to sum the derivative expansion in QED using other methods \cite{Gusynin,Gusynin_2}. Also, we would like to thank A. Zelnikov for pointing out to us references that also deal with the problem of derivative summation \cite{Barvinsky,Barvinsky_2}. 

% If you have acknowledgments this puts in the proper section
\begin{acknowledgments}
Work partially supported by the CICYT Research Project
AEN99-0766.
% put your acknowledgments here.
\end{acknowledgments}

% Create the reference section using BibTeX
%\bibliography{your bib file}

% figures follow here or may be put into the text as floats.
% Use the graphics or
% graphicx packages distributed with LaTeX2e. See the LaTeX Graphics
% Companion by Michel Goosens, Sebastian Rahtz, and Frank Mittelbach
% for instance.
%
% Here is an example of the general form of a figure:
% Fill in the caption in the braces of the \caption{} command. Put the label
% that you will use with \ref{} command in the braces of the \label{} command.
%
 
% tables follow here or maybe be put in the text
%
% Here is an example of the general form of a table:
% Fill in the caption in the braces of the \caption{} command. Put the label
% that you will use with \ref{} command in the braces of the \label{} command.
% Insert the column specifiers (l, r, c, d, etc.) in the empty braces of the
% \begin{tabular}{} command.
%
% \begin{table}
% \label{}
% \caption{}
% \begin{tabular}{}
% \end{tabular}
% \end{table}

\end{document}